# Real-World Choreographic Programming: Full-Duplex Asynchrony and Interoperability

Lovro Lugović[a]  and Fabrizio Montesi[a]

a    Department of Mathematics and Computer Science, University of Southern Denmark, Denmark

**Abstract**    In the paradigm of *choreographic programming*, the overall behaviour of a distributed system is coded as a *choreography* from a global viewpoint. The choreography can then be automatically *projected* (compiled) to a correct implementation for each participant. This paradigm is interesting because it relieves the programmer from manually writing the separate send and receive actions performed by participants, which simplifies development and avoids communication mismatches.

However, the applicability of choreographic programming in the real world remains largely unexplored. The reason is twofold. First, while there have been several proposals of choreographic programming languages, none of these languages have been used to implement a realistic, widely-used protocol. Thus there is a lack of experience on how realistic choreographic programs are structured and on the relevance of the different features explored in theoretical models. Second, applications of choreographic programming shown so far are intrusive, in the sense that each participant must use exactly the code projected from the choreography. This prevents using the code generated from choreographies with existing third-party implementations of some participants, something that is very beneficial for testing or might even come as a requirement.

This paper addresses both problems. In particular, we carry out the first development in choreographic programming of a widespread real-world protocol: the Internet Relay Chat (IRC) client–server protocol [28]. The development is based on Choral, an object-oriented higher-order choreographic programming language (choreographies can be parametric on choreographies and carry state).

We find that two of Choral's features are key to our implementation: higher-order choreographies are used for modelling the complex interaction patterns that arise due to IRC's asynchronous nature, while user-definable communication semantics are relevant for achieving interoperability with third-party implementations. In the process we also discover a missing piece: the capability of statically detecting that choices on alternative distributed behaviours are appropriately communicated by means of message types, for which we extend the Choral compiler with an elegant solution based on subtyping.

Our Choral implementation of IRC arguably represents a milestone for choreographic programming, since it is the first empirical evidence that the paradigm can be used to faithfully codify protocols found 'in the wild'. We observe that the choreographic approach reduces the interaction complexity of our program, compared to the traditional approach of writing separate send and receive actions. To check that our implementation is indeed interoperable with third-party software, we test it against publicly available conformance tests for IRC and some of the most popular IRC client and server software. We also evaluate the performance and scalability of our implementation by performing performance tests.

Our experience shows that even if choreographic programming is still in its early life, it can already be applied to a realistic setting.

## The Art, Science, and Engineering of Programming

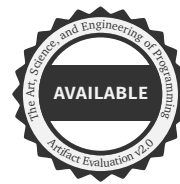 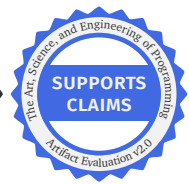

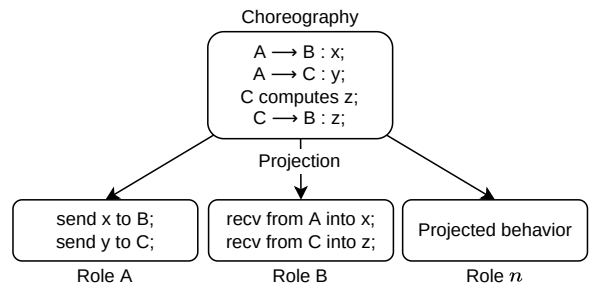

**Figure 1** Choreographic programming (illustrated with the Alice and Bob notation).

# 1 Introduction

**Choreographic programming**  *Choreographic programming* [36] is a recent programming paradigm, whose goal is to simplify the implementation of concurrent and distributed software. Unlike in traditional development, where the programmer produces a separate program for each *role* (a participant) within the distributed system, choreographic programming allows for writing the collective behaviour of these roles and the way they interact from a 'global viewpoint' in an artifact called a *choreography*. The choreography can then be mechanically transformed by a compiler into executable code for each role, a process called *endpoint projection (EPP)* [2, 5, 6, 21].

We illustrate choreographic programming using pseudocode in Figure 1. Essentially, a choreography expresses directly the interactions that should take place among the roles of interest, which are then compiled to the necessary send and receive actions to make these interactions happen at runtime.

Thus, in addition to providing an explicit description of the interactions that should take place, the paradigm lifts the burden of having to manually match the necessary send/receive communication actions. Theories of choreographic programming come with strong safety and liveness guarantees based on this aspect, like lack of communication errors (two roles never try to perform incompatible actions directed at each other) and deadlock-freedom [6, 21]. The simplicity and advantages offered by choreographic programming motivated the implementation of several prototypes [6, 15, 19, 41], the development of many theoretical models and extensions [7, 13, 27, 32, 33, 34], and the mechanisation of its foundations in theorem provers [14, 23, 41]. An introduction to the foundations of the paradigm is given in [37].

**The problem**  Despite the rising interest in choreographic programming, there is still a considerable lack of practical validation. The paradigm's potential and applicability in the real world remain largely unexplored. In particular, to date choreographic programming has yet to be used in the implementation of a realistic, widely-used protocol. Such an activity would hopefully provide insight into how choreographic programs can codify the informally-specified interaction patterns found in the wild.

**This work**   We carry out the first development of a widespread real-world protocol using choreographic programming. Specifically, we program the Internet Relay Chat (IRC) client–server protocol [26, 28] in the choreographic language Choral [19].

IRC is an ideal candidate for our study because it is a mature protocol that has been widely used for decades. This gives us an environment that is well-understood, stable, and rich with third-party implementations. IRC also strikes the right balance between simplicity and interesting concurrent behaviour for a first realistic exploration of choreographic programming. The protocol comprises two roles – the minimum to have a meaningful application of choreographies – and exhibits advanced nondeterministic behaviour: the client and the server can simultaneously send messages to each other without having to wait for replies to previous messages, following a pattern that we call 'full-duplex asynchronous communication'. Capturing this kind of nondeterministic communication behaviour in choreographies is known to be challenging [37], and this pattern remains uncharted territory.

Choral was chosen as it is arguably the most mature choreographic programming language currently available. We found its support for higher-order choreographies (choreographies parameterised on choreographies) particularly useful to capture and generalise the complex interaction patterns that we identified within IRC. Furthermore, Choral was extensible enough for us to define a communication semantics that complies with IRC's wire format.

Our key contributions are:

- A choreographic implementation of IRC, given in terms of a Choral program (Sections 3 and 4). Our implementation can interoperate with several of the most popular third-party implementations of IRC clients and servers (Section 5).

- A modular choreographic library that, for the first time, captures full-duplex asynchronous communications (Section 4.1). Capturing this pattern is necessary to implement IRC. Our library is designed to be reusable in the future.

- An extension of the Choral compiler to *type-driven selections*, a new mechanism that is key to interoperability (Section 5.2). When the behaviour of a role depends on a choice made by another role, choreographic compilers must check that this choice is correctly communicated [8, 37]. Our mechanism introduces the flexibility to communicate choices through user-definable parts of message payloads.

- An evaluation of our implementation, including conformance and performance tests (Section 6).

We also discuss some of the advantages and challenges encountered in our application of choreographic programming, which point to useful future improvements of choreographic programming languages (Section 7).

## 2   Background

We begin with an overview of the relevant background on the Choral programming language [19] and the IRC protocol specification [26, 28].

## 2.1 Choreographic Programming in Choral

Choral [19] is an object-oriented choreographic programming language. Its syntax closely mirrors that of Java, but extends all types and literals to express the roles at which they are located. What makes Choral choreographic is that a type might mention multiple roles, denoting distribution. For example, a user-defined class `MyClass@(A, B, C)` is a choreography whose code can involve the roles `A`, `B` and `C`, and communications between them. Here `A`, `B` and `C` are formally role variables bound within the definition of `MyClass`, which must be instantiated when the type is used. Choral code can then be projected to separate Java code for each role, in this case three Java classes called `MyClass_A`, `MyClass_B`, and `MyClass_C`, each containing the respective local behaviour that the roles should run together to implement the original choreographic code. The object-oriented aspect stems from choreographies being objects: they can be instantiated many times, passed around as parameters and returned from methods, naturally leading to *higher-order choreographies* where one is parameterised over others [11, 12, 16, 19, 23].

Roles in Choral programs can communicate through *channels*, which are objects with methods that move data of some type from one role to another. Choral has a hierarchy of channel interfaces with different properties: the direction of communication, the type of the data, etc. The next snippet shows two of the more common channel interfaces: `DiChannel@(A, B)<T>` (a one-directional channel) and `SymChannel@(A, B)<T>` (a bidirectional channel for communicating data of the same type in both directions).

```
public interface DiChannel@(A, B)<T@X> {
  <S@Y extends T@Y> S@B com(S@A m);
  @SelectionMethod <T@X extends Enum@X<T>> T@B select(T@A m);
}

public interface SymChannel@(A, B)<T@X> extends DiChannel@(A, B)<T>,
                                                 DiChannel@(B, A)<T> {}
```
*Choral*

Note how method `com` takes data of type `T` at role `A` and returns data at role `B`, capturing communication. Method `select` is a specialisation of `com` that supports only enumeration types and uses the special `@SelectionMethod` annotation to instruct the Choral compiler that its invocations represent *selections*: communications of choices. Selections are used to address a problem known as *knowledge of choice* [8, 37]. We demonstrate this problem with a choreography where role `A` transfers an integer sequence of unknown (but finite) size to role `B`.

```
enum Choice@A { GO, STOP }

class Transfer@(A, B) {
  private SymChannel@(A, B)<Integer> ch;
  public Transfer(SymChannel@(A, B)<Integer> ch) { this.ch = ch; }

  void consumeItems(Iterator@A<Integer> it, Consumer@B<Integer> f) {
    if (it.hasNext()) {
      ch.<Choice>select(Choice@A.GO);
      f.accept(ch.<Integer>com(it.next()));
      consumeItems(it, f);
    } else {
      ch.<Choice>select(Choice@A.STOP);
    }
  }
}
```
*Choral*

Method `consumeItems` uses selections to communicate the knowledge of whether the sequence is empty or not. If A determines that the sequence is non-empty, it tells B to expect a value by `select`ing GO and communicates the integer to be consumed, after which they both recur to transfer the remaining items. Otherwise, A tells B that there is nothing left to do by `select`ing STOP and they both return from the method.

Projecting the choreography results in the distributed code shown below: there is one program per role (`Transfer_A` and `Transfer_B`), and each program contains only the code that pertains to that role.

```
class Transfer_A {
  private SymChannel_A<Integer> ch; ...
  void consumeItems(Iterator<Integer> it){
    if (it.hasNext()) {
      ch.<Choice>select(Choice.GO);
      ch.<Integer>com(it.next());
      consumeItems(it);
    } else {
      ch.<Choice>select(Choice.STOP);
    }
  }
}
```
*Projection*

```
class Transfer_B {
  private SymChannel_B<Integer> ch; ...
  void consumeItems(Consumer<Integer> f) {
    switch (ch.<Choice>select()) {
    case GO -> {
      f.accept(ch.<Integer>com());
      consumeItems(f);
    }
    case STOP -> { /* done */ }
    default -> { /* raise error */ }
    }
  }
}
```
*Projection*

Both projections of the method have just one parameter, as projection discards any parameters whose type does not involve the role being projected. A makes a choice by evaluating the boolean guard, so the conditional is exactly present in its projection (but not B's). It also invokes the single-parameter projections of `com` and `select` to perform its send actions.

The code for B on the other hand uses the parameterless projections of `com` and `select` for its receiving actions. It starts off by receiving a selection label from A that tells it what to do: either receive and recur in case of GO or terminate in case of STOP. The **switch** block is automatically generated by the Choral compiler and is what makes B react to the external choice made by A. Choral checks for knowledge of choice statically and rejects code where roles affected by external choices do not have enough information to discern them. For example, removing the STOP selection in the choreography would make it unprojectable because B would not know when to terminate. In the Choral compiler, choices always have to be explicitly communicated via selections of dedicated enumerated values [19]. As we are going to see, this is not expressive enough to implement IRC, which will prompt the introduction of type-driven selections in Section 5.2.

As we mentioned in Section 1, Choral, and choreographic programming in general, bring some notable benefits [37]. Choreographies are shorter than their implementations (which can lead to fewer bugs [4]). The communication actions executed by roles are always compatible and communicate messages of the right types (communication safety). The compiler only accepts code that guarantees knowledge of choice. Under the assumption that internal computations terminate correctly, the generated code is deadlock-free. Furthermore, in this work (Section 7), we are going to elicit an additional advantage: the complexity of how software components interact is lower at the choreographic level.

## 2.2 Internet Relay Chat (IRC)

Internet Relay Chat (IRC) [26] is a text-based chat protocol that originated in 1988. Today, it remains widespread: the top 100 IRC networks serve approximately quarter to half a million users at a time, and are especially popular in the free and open-source software (FOSS) community.

The core of IRC is a TCP-based client–server protocol [28] that establishes how clients can communicate using *channels* (i.e., 'rooms', not to be confused with Choral channels) and *private messages*. Upon connecting, users can choose a *nickname* and can join and leave channels at will. Channels are created dynamically and serve to group users, with a message to a channel being broadcast to all of its members.

Clients and servers communicate *IRC messages* to each other, which have a uniform format. On the wire, IRC messages are (most often) UTF-8 encoded strings that consist of a *command* and zero or more space-separated *parameters*, with the last parameter optionally being preceded by a colon to allow for the inclusion of spaces: `COMMAND param1 param2 :trailing param`. Messages sent by the server must also contain a prefix called the *source*, which starts with a colon and indicates the originator of the message, usually either the server's hostname or a client's nickname. For example, the IRC message `:bob PRIVMSG #compsci :Hello there` would be sent by the server to members of the `#compsci` channel to notify them that `bob` sent `Hello there` as a message to the channel. For brevity, we often omit sources in our examples.

IRC defines a number of commands, whose meaning depends on the side that sends them. A special class of commands sent only by the server as replies to clients' requests are called *numeric replies*, since their names consist only of digits. They are given symbolic names by the specification for convenience, but only their numeric names can ever be used on the wire. Table 1 shows some of the common IRC commands.

Although most requests generate some sort of reply from the other participant, the number of reply messages is not fixed (zero or more), as shown in Figure 2.

■ **Table 1** Some common IRC commands, along with short descriptions of their meanings depending on the direction of the communication.

| Command | Client → Server | Server → Client |
|---|---|---|
| PING | ping the server | ping the client |
| PONG | reply to a ping | reply to a ping |
| NICK | set the nickname | acknowledge the nickname |
| USER | set the username | n/a |
| JOIN | join a channel | acknowledge or announce a join |
| PRIVMSG | send a message | deliver a message |
| 001 (RPL_WELCOME) | n/a | registration complete |

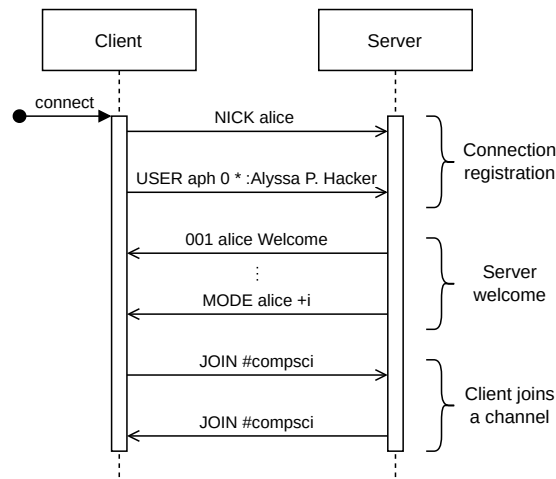

■ **Figure 2** An example IRC conversation. The NICK–USER sequence is the most basic form of *connection registration* initiated by the client, after which the server sends a welcome and the client is free to send other commands. Notice that the number of reply messages to a command is variable, ranging from zero to many.

## 3 Architecture

In this section we present the overall architecture of our implementation, before showing its details in Section 4.

We model IRC as the two-role choreography Irc@(Client, Server) which projects to two Java classes, Irc_Client and Irc_Server. While the choreography only models the interactions between a single client and server, in reality the server has to be able to support many clients at the same time. The way we achieve this and the overall architecture of our implementation are shown in Figure 3. Specifically, the code projected for the server (Irc_Server) is instantiated once for each client that connects, and all instances are run concurrently. These instances access a state component to manage global information, like the set of connected clients, the set of currently existing channels and their members, etc. It is important to note that our development does not fix the implementation of this component. While we offer a default based on an in-memory shared object, in general this component can be replaced with any

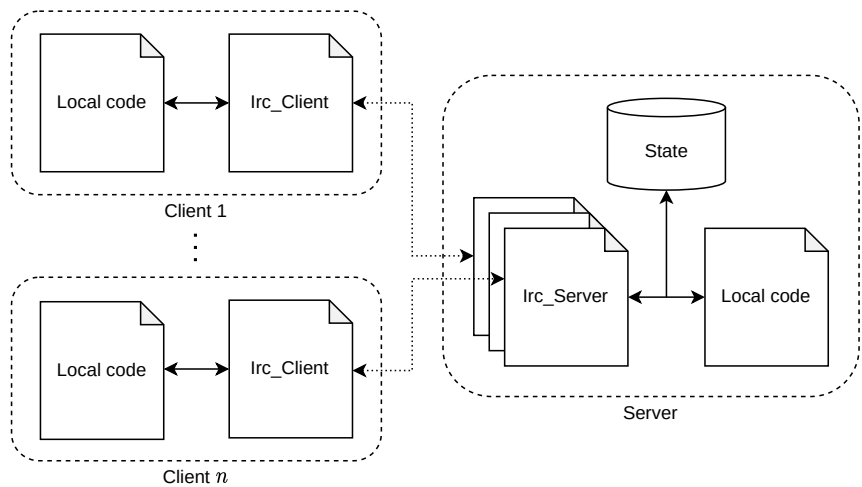

■ **Figure 3** Architecture of the choreographic IRC implementation.

implementation that respects the interface. For example, a distributed database could be used to support replicating the server, without any changes to the choreography.

The implementation of the client on the other hand instantiates its projected code (`Irc_Client`) only once. Implementations of both the server and the client contain pieces of *local code* written in Java that are used to provide a simple user interface and call into the Java libraries projected from the choreography. For choreographies written in a modular way such as ours, this mechanism can be used to *incrementally* adopt choreographic programming within an existing codebase capable of interacting with Java libraries.

## 4 The Choreography

### 4.1 Full-duplex Asynchronous Communication

Most of the complexity in programming the IRC choreography lies in the *full-duplex asynchronous* nature of the protocol mentioned in Section 1: not only can the client and the server send (and receive) messages simultaneously without having to wait for each other (full-duplex), but responses from either side can arrive out-of-order compared to their initial requests (asynchronous). Figure 4 illustrates one such case.

We first show how we capture full-duplex asynchrony by using an event queue model, which is the heart of our IRC choreography in Section 4.2. Specifically, we employ a pair of queues whose *events* are processed by a corresponding pair of concurrently-running choreographic event *handlers*, as shown by Figure 5 (left side). Observe that the pattern shown in the figure is completely symmetric and not at all specific to IRC, making it suitable for reuse by choreographies of other protocols with similar concurrent behaviour. This is another reason for why IRC is a good candidate for our exploration of choreographic programming in practice. For the time being, we keep things general and refer to the two roles as just A and B, rather than Client

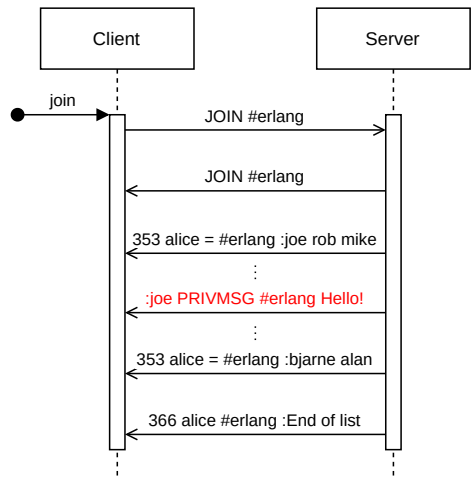

■ **Figure 4** An example of the asynchronous nature of the IRC protocol. The client receives a channel's members after requesting to join the channel, but can at any moment receive any other IRC message, such as `PRIVMSG`.

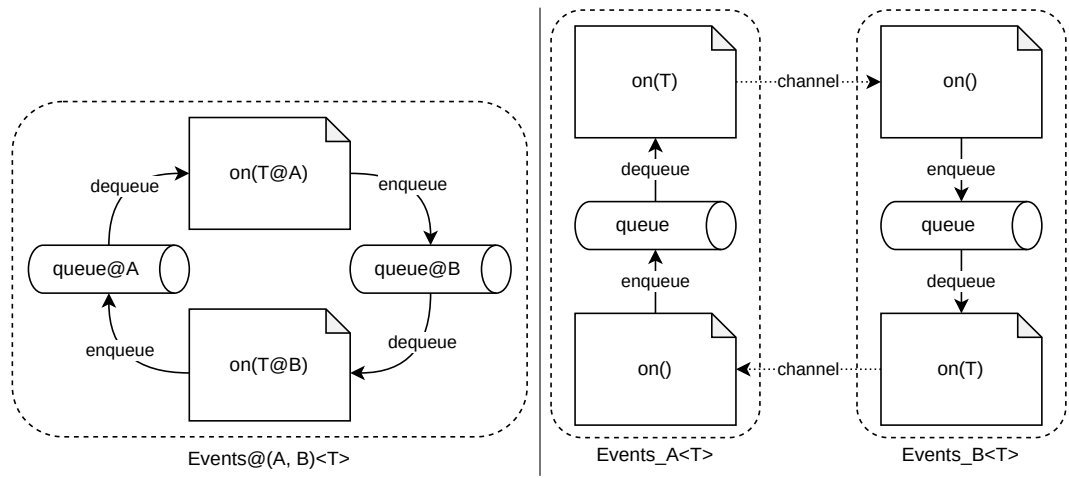

■ **Figure 5** Left: Architecture of the full-duplex asynchronous choreographic pattern; events are stored within two queues, managed by two concurrent event handlers represented by the methods `on(T@A)` and `on(T@B)`. Right: The two projections of the pattern. Notice the two complementary directions of communication.

and `Server`. In our setting, the events will be the IRC messages exchanged by the two participants, but could in general be any sort of data.

Each queue contains the events that should be processed by a specific role. The event queues are manipulated by two event loops running concurrently in the background (not shown in the diagram). A loop takes care of handing off events from the corresponding queue to the correct handler, `on(T@A)` or `on(T@B)`. While the event queues and the events within them are local to each role, the event handlers are choreographic code that involve both roles. This means that when a role processes an event it can decide to communicate data to the other side. The receiver can then do some processing of its own based on the message, and as a result might decide to enqueue more events onto its own queue for later handling. The interplay between the two event handlers is what drives the system.

The projections of the choreography for each role are shown in Figure 5 (right side). Each contains a single event queue, but also includes the projections of *both* of the event handlers, since both roles are involved in each of them. The event handler projections run concurrently, which is what enables the two participants to send and receive messages at the same time. At A, the projection on(T) of handler on(T@A) is the one handling events from the queue and communicating information to B, while the projection on( ) of on(T@B) is the one receiving information from B and enqueuing new events (vice versa for the projections at B).

If the communication channel supports it, it is possible for the event handlers to share a single full-duplex channel and use it concurrently, each in its own direction. This is the case of IRC, which uses TCP as transport. While the two directions of a full-duplex channel can be used concurrently, concurrent usage of one direction must be handled carefully. In particular, if a role performs two concurrent sends and the other two concurrent receives, the messages might be received in the wrong order. Writing such code would be a programming mistake and it is therefore important that the handlers are in charge of opposing communication directions over the channel. Our implementation statically enforces this constraint by leveraging Choral's type system, and in particular by using Choral's directed (one-way) channel interface DiChannel. A DiChannel@(A, B)<T> can be made available to on(T@A), and a DiChannel@(B, A)<T> to on(T@B), ensuring that the communication directions of the two handlers are different. These two directed channels are instantiated by the caller from the same bidirectional channel. Aside from the concurrency considerations above, there is otherwise no particular constraint on the kind of channels that can be used with our choreographic pattern.

**The Events library** To lift the burden of arranging for the projections to run concurrently and allow for reuse of the pattern in choreographies that involve similar asynchronous interactions, we packaged the implementation of the pattern into a library offering a higher-order choreography called Events@(A, B), shown below.

```
public interface Events@(A, B)<T@X> {
  void run(EventHandler@(A, B)<T> eventHandlerA,
           EventHandler@(B, A)<T> eventHandlerB,
           LocalHandler@A localHandlerA,
           LocalHandler@B localHandlerB);
  EventQueue@A<T> queueA();
  EventQueue@B<T> queueB();
}
public interface EventHandler@(A, B)<T@X> { void on(T@A event); }
public interface LocalHandler@X {
  boolean@X onError(Exception@X e);
  void onStop();
}
public interface EventQueue@X<T@X> { void enqueue(T@X event); }
```
*Choral*

The choreography is parametric with respect to the kind of events being communicated, taking the event type as a type parameter. For simplicity we only accept a single type parameter, but we could have also chosen to use two in order to allow the two roles to have different event types. The library abstracts away the low-level details of managing the concurrently-running handlers and the event queue, while exposing enough functionality and hooks so that the main choreography can define how the events are handled and generated.

`Events` provides three methods, with `run` as the entrypoint. Its two parameters of type `EventHandler` expose the mentioned `on` methods for event handling. The location of an event received by an `on` method depends on the the first role parameter of `EventHandler`, which is why the two handler parameters have reversed role lists. Calling `run` will start the concurrent execution of `eventHandlerA`'s and `eventHandlerB`'s `on` methods (both blocking on their respective queues until an event is ready) and block the caller until the choreography terminates. Methods `queueA` and `queueB` can be used to get an instance of `EventQueue`, which only allows for enqueueing new events (a thread-safe operation).

**Local events and error handling**　The two single-role parameters of type `LocalHandler` are provided as hooks for the local code to react to errors or the event loop stopping (e.g., due to a `QUIT` message). In particular, method `onError` is used to handle exceptions thrown within either of the two event handlers. The method receives the exception thrown and has to decide whether to terminate the choreography by returning an appropriate Boolean value. This is the primary way in which we handle faults and detect situations such as clients disconnecting ungracefully by terminating the connection, etc. Method `onStop` is provided as a hook for the local code to run any necessary cleanup once the choreography has fully terminated.

As far as we know, `Events` is the first case of a reusable library for a nontrivial choreographic pattern. Its definition is based on higher-order composition, which was only recently introduced to choreographic programming with the Choral language [19].

## 4.2 Internet Relay Chat

**Main choreography and messages**　Armed with our choreographic pattern and new library, we can now examine how the top-level `Irc` choreography is constructed. In the context of IRC, the events that the participants process are IRC messages. A message is represented by a `Message` object that provides methods for accessing its components and (de)serialising it to/from a string.

```
enum Command@X { NICK, USER, JOIN, PRIVMSG, ... }

class Message@X {
  ...
  public static Message@X parse(String@X str) { ... }
  public String@X toString() { ... }
  public Command@X getCommand() { ... }
}
```
*Choral*

Implementing IRC now boils down to using the `Events` pattern and providing the appropriate definitions for its event handlers, which we do below.

```
class Irc@(Client, Server) {
  private Events@(Client, Server)<Message> events; ...

  public void run() {
    events.run(
      new IrcClientToServerHandler@(Client, Server)(
        events.queueB(), ch, cState, sState, cId),
      new IrcServerToClientHandler@(Client, Server)(
        events.queueA(), ch, cState, sState, cId),
      new IrcClientLocalHandler@Client(events.queueA(), cState),
      new IrcServerLocalHandler@Server(events.queueB(), sState, cId));
  }

  public void enqueue(Message@Client m) { events.queueA().enqueue(m); }
  public void enqueue(Message@Server m) { events.queueB().enqueue(m); }
}
```
*Choral*

The choreography provides a `run` method of its own that invokes `Events` with the necessary arguments (other choreographies). All of the choreographies get references to their required components: the event queues, the channel for communicating (`ch`), an object to interact with the client's state and user interface (`cState`), and the necessary server state to keep track of the participants' details (`sState`, `cId`). The IRC interface also exposes the pattern's `enqueue` methods that can be used to add new events into the system. That is the local code's primary way of interacting with the choreography. Running the choreography by invoking the `run` method only sets up the concurrent execution of the event handlers that block waiting for events from the queues. The local code must then kick-start the interaction by enqueueing new events. At the client this might happen as a consequence of a user interaction, while at the server it might be an automated job, such as periodically pinging the clients.

**The handlers**    In our implementation we model each kind of IRC command as a subtype of `Message` (we maintain a strict one-to-one mapping between the two), e.g. `PingMessage`, `UserMessage`. This provides the user of the choreography with a more semantically rich interface, but also gives us a technical benefit that allows us to deal with interoperability, as we are about to describe.

When an event is to be processed, it is delivered to the corresponding handler as a `Message` object located at a particular role. The role then picks an appropriate code branch based on the message's type. Because the handlers are choreographic code, this knowledge is initially known only to that role and must be explicitly communicated to the other one by means of a selection when the branch is taken. Thus Choral requires two communications for each IRC message: one for the selection, and the other for the actual data. This is not compatible with IRC, which prescribes only one communication: the separate selection is not needed because the choice is already captured by the command that is included in the message together with its parameters.

We capture this style of communicating knowledge of choice as part of a message payload with *type-driven selection*. We postpone the details of the mechanism until

Section 5.2 and for now describe only the parts relevant for the handlers that we are about to show. At the API level, our mechanism consists of an extension to Choral's `DiChannel` interface with a `tselect` method:

```
public interface DiChannel@(A, B)<T@X> {
  @TypeSelectionMethod public <S@X extends T@X> S@B tselect(S@A m); ...
}
```
*Choral*

The method is a combination of the previously-shown `com` and `select` – it moves a value of any subtype `S` of `T` from `A` to `B`, but it is also treated as a selection by the Choral compiler due to the new `@TypeSelectionMethod` annotation. Essentially, the type `S` of the communicated value is used in place of an explicit selection label. The next snippet shows the structure of the handler for events from the `Client`'s queue.

```
class IrcClientToServerHandler@(Client, Server)
    implements EventHandler@(Client, Server)<Message> {
  ...
  public void on(Message@Client m) {
    switch (m.getCommand()) {
    case PING -> {
      PingMessage@Server ping = ch.<PingMessage>tselect((PingMessage)m);
      if (ping.hasEnoughParams())
        sQueue.enqueue(new PongMessage@Server(...));
    }
    case USER -> {
      UserMessage@Server user = ch.<UserMessage>tselect((UserMessage)m);
      if (!State.isRegistered(cId) && ...)
        sQueue.enqueue(new WelcomeMessage@Server(...)); ...
    } ...
    }
  }
}
```
*Choral*

Depending on the command of the message, a corresponding branch of the **switch** is chosen. Choral does not yet provide an **instanceof** operator so we recover the concrete type with a manual cast after inspecting the command. (Casts are not part of Choral either, but we use them as syntax sugar for a utility function that we wrote in Java.) Each branch then uses the new `tselect` method to both communicate the message *and* make the Choral compiler aware that a selection is taking place. The two projections of the handler show the type-driven selection in action more clearly:

```
class IrcClientToServerHandler_Client
  implements EventHandler_A<Message> {
  public void on(Message m) {
    switch (m.getCommand()) {
    case PING -> {
      ch.<PingMessage>tselect((PingMessage)m);
    }
    case USER -> {
      ch.<UserMessage>tselect((UserMessage)m);
    } ...
    }
  }
}
```
*Projection*

```
class IrcClientToServerHandler_Server
  implements EventHandler_B<Message> {
  public void on() {
    switch (ch.<Message>tselect()) {
    case PingMessage ping -> {
      if(ping.hasEnoughParams())...
    }
    case UserMessage user -> {
      if(!State.isRegistered(cId))...
    } ...
    }
  }
}
```
*Projection*

On the sender's side the structure of the projection is like the one in the example in Section 2.1, except that calls to `com` and `select` have been merged into the single `tselect` call. The receiver, instead, first receives an object of the generic `Message` type, and then uses its dynamic type to select the appropriate branch (via a *pattern switch expression*). Compared to the receiver in the example in Section 2.1, the dynamic type of the received message now plays the role of the selection label. The nested conditionals are local to the server so a selection is not necessary in their case.

The code for handling events from the `Server` queue has a similar structure:

```
class IrcServerToClientHandler@(Client, Server)
    implements EventHandler@(Server, Client)<Message> {
  ...
  public void on(Message@Server m) {
    switch (message.getCommand()) {
    case RPL_WELCOME -> {
      WelcomeMessage@Client w = ch.<WelcomeMessage>tselect((WelcomeMessage)m);
      cState.println("Welcome: "@Client + w.toString()); ...
    } ...
    }
  }
}
```
Choral

To illustrate the interaction between the handlers, consider the case of `Client` registering the connection with `Server`, as in Figure 2. The operation starts externally with NICK and USER messages being enqueued onto the client's queue (e.g., by the local code driving the user interface). Both messages are picked up by the `on(Message@Client)` method of `IrcClientToServerHandler`, which chooses the corresponding case and communicates the message via a type-driven selection. Upon reception, the server performs its own local checks and if the messages are invalid it enqueues an appropriate error reply. Otherwise, if the messages are valid and both have been received from the client, the connection registration sequence is complete and the server enqueues a series of welcome messages (RPL_WELCOME, etc.). Each of these is handled by the `on(Message@Server)` method of `IrcServerToClientHandler`. When a client receives them, it can update its state about the server and display information to the user. The implementation uses the same pattern to implement handling of all other message types necessary for a working IRC implementation.

## 5   Interoperability

### 5.1  Controlling the Wire Format

Respecting the wire format dictated by IRC requires careful manual control. In our case, this corresponds to determining how our `Message` objects are (de)serialised by channels. Recalling Section 2.2, communicating an object of type `PingMessage` should actually end up transferring the UTF-8 encoded string PING <token>.

To address this issue we make use of Choral's extensibility and implement a custom Choral channel – `IrcChannel@(A, B)`. It implements Choral's `SymChannel` interface (see Section 2.1), which makes it a subtype of both `DiChannel@(A, B)` and `DiChannel@(B, A)`. Because the channel is purpose-specific, we restrict the set of values it is able to transfer to just subtypes of our `Message` type. The implementation is conceptually simple and implements the (de)serialisation of `Message` according to IRC's rules. The class shown below is the base of the implementation, which is essentially a wrapper around Java's `ByteChannel` that contains methods used by both projections of `IrcChannel`. We omit the helper methods and the less interesting details for the sake of clarity (buffering of the messages, etc.).

```
class IrcChannelImpl implements SymChannelImpl<Message> {
  ...
  public <M extends Message> Unit tselect(M m) {
    write(socket, encode(m.toString()));
  }
  public <M extends Message> M tselect() {
    read(socket, buffer, MAX_SIZE);
    Message m = Message.parse(extractUntil(buffer, CRLF));
    ... // Verify that parsing succeeded and create the proper subtype.
    return (M)m;
  }
}
```
<div align="right">Local</div>

The two `tselect` methods implement the conversion between a subtype of `Message` and a sequence of bytes sent across the wire. To send a message, `tselect(M)` calls `m.toString()` to generate an IRC message string and sends it over the channel. Since we are dealing with a TCP stream, the method `tselect()` for receiving a message has to take care of buffering its input and detecting the message boundaries. Once a message has been picked out it can be parsed and an appropriate subtype of `Message` can be constructed based on its command. In case invalid data is received, the channel will throw an exception that can be handled by the local code.

## 5.2 Type-driven Selections

Another big aspect of interoperability is communicating choices as part of message payloads. We introduced this in Section 4 and now describe the remaining details.

To implement type-driven selections, we had to extend the projection procedure of Choral's compiler. The starting point is the addition of the `@TypeSelectionMethod` annotation, which instructs the compiler to treat calls to the annotated method as type-driven selections. We then exploit Choral's type system and its subtyping relation inherited from Java. The key idea is to use the (static and dynamic) type of the communicated value instead of a selection label, as illustrated in Section 4 by `IrcClientToServerHandler_Server`. Specifically, we distinguish branches by using different subtypes of the common supertype of messages that can be communicated along the channel. The static type gives the Choral compiler sufficient information to perform its usual correctness checks and project the code.

At the receiver's side, the dynamic type of the deserialised object received through the channel is used to select the appropriate branch and recover the static type information known to the Choral compiler at the time of projection. In order to ensure the correctness of this approach and project the code properly, we had to extend the Choral compiler to check for two things.

First, the types of messages communicated by type-driven selections in different branches need to have a *common supertype*, which itself is a subtype of what the channel can communicate. This will be the initial type of the message at the receiver, before its dynamic type is inspected in order to make the choice. The supertype is guaranteed to exist if the choreography is well-typed, which is checked by Choral's type system. Our extension of the Choral compiler finds the most precise such supertype.

Second, the same message types have to be *disjoint*: none is a subtype of another and they are all class types. This property ensures that the approach is sound – for each message, there is exactly one branch of the projected `switch` at the receiver that can be entered. If we had subtyping overlaps between the different branches, the choice would be ambiguous and inspecting the message's dynamic type would not be enough to guarantee uniqueness. The class type requirement is to avoid ambiguity because of the implementation of multiple interfaces. Additionally, the disjointness check is performed against the erased versions of the types, as Java's generic type parameters cannot be inspected at runtime.

## 6 Evaluation

### 6.1 Conformance Tests

To test that our implementation is indeed interoperable, we used the code projected for the `Client` and the `Server` in combination with some of the most popular IRC servers and clients available today, respectively. Both projections generated from our choreography were able to perform all of the basic functionalities required for a normal chat session between multiple users.

Specifically, the code compiled for `Client` was tested with the popular Libera Chat [10] and W3C [1] IRC networks. This consisted of connecting multiple concurrent instances of the client to each network, configuring their nicknames, joining the same channel, sending messages between the clients individually as well as to the channel, and confirming that all parties received each other's messages.

On the other hand, the code for `Server` was tested with the WeeChat [22], Irssi [42], HexChat [9] and Konversation [29] IRC clients. For each of the third-party clients we performed the same sequence of steps as above, checking that our server was successful in accepting the clients' connections and delivering their messages.

We have also tested our server implementation with a publicly available suite of IRC conformance tests called `irctest` [35]. The suite uses a Python driver to connect to a specified server and test whether is is able to accept and reply to standard IRC messages in a conforming way. Our server implementation passes all of the tests related to the basic functionalities described above, in particular, the tests for the `NICK`, `USER`, `JOIN`, `PART`, `PRIVMSG`, `QUIT`, `PING` and `PONG` messages. The full invocation of the suite and a list of the passed tests are given in Appendix A.

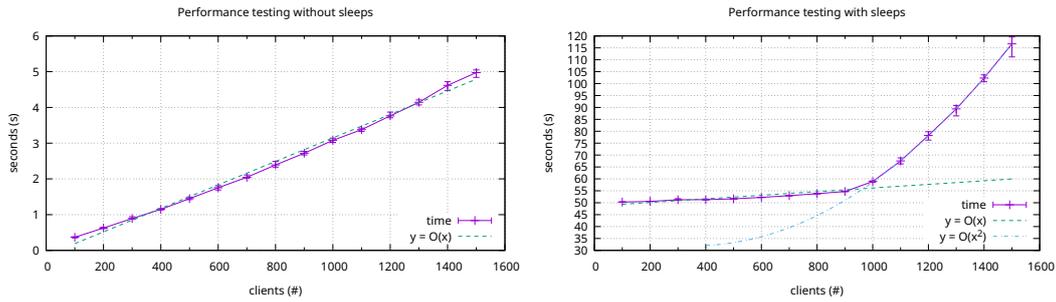

**Figure 6** Wall-clock execution times when clients do not (left plot) or do (right plot) significantly interact through IRC channels.

We did not implement any other features of the IRC client–server protocol, such as user and channel modes or capability negotiation, as they are not necessary to demonstrate interoperability. However, the choreography could straightforwardly be extended to support additional messages, and we believe that this does not present any new conceptual challenges with regards to our work.

### 6.2 Performance Tests

Although performance was not the primary goal of our development, we carried out tests to check the scalability of the IRC server implementation generated from our choreography. We performed two types of performance tests on an Arch Linux system (Linux 6.1.6) with an AMD Ryzen 9 3950X processor and 64 GiB of main memory. Both tests simulate up to 1500 concurrent clients connecting to the server. For the purposes of these tests, we implemented the clients as shell scripts using standard command-line tools to keep the resource usage low. When connected, each client joins a predefined channel and sends three messages which get broadcast to all of the other members. Once done, the client sends a quit message and disconnects.

The first type of performance test simulates a scenario where clients connect to the server and execute their actions as soon as they can. Figure 6 (left side) displays a plot of the measurements, with the axes representing the number of clients and the wall-clock execution time (in seconds) of the test. We ran multiple tests in succession with a linearly increasing number of clients. The solid purple line plots the execution time of each test averaged over 5 runs, while the dashed green line is a linear curve fit to the data. The graph shows that the execution time scales linearly with the number of clients. This indicates that creating and destroying instances of (the projection of) the choreography (one for each client) performs well, at least in the scope of our test.

The second type of performance test follows the same pattern as the first, except that clients now perform a 10 second pause ('sleep') after connecting to the server and an additional 2 second pause between each sending action. The pauses create a situation where every client is connected to the server and can see the other clients present in the channel. Since a message sent to a channel is broadcast to all of its members, this means that the server will end up performing approximately $3n^2$ message exchanges where $n$ is the number of clients. Figure 6 (right side) displays the measured results.

As before, we ran multiple tests in succession and plotted the execution time of each test averaged over 5 runs. The dashed green and dash-dotted cyan lines are linear and quadratic curves fit to portions of the data to illustrate the scalability of the server. The execution time scales linearly up to roughly 1000 clients, indicating that until then it benefits considerably from performing communications in parallel. After that threshold our implementation starts to scale quadratically. This is likely because we use a native-thread-per-client concurrency strategy for the sake of simplicity. When the number of context switches becomes too large, the benefits of parallelism diminish and performance starts correlating with the number of exchanged messages. A direct way to improve performance would be switching to Java's recent 'virtual threads'.

Overall, this preliminary performance evaluation indicates that our prototype implementation works well, considering that we paid no attention to optimisation. We leave performance improvements to future work as our measurements are adequate to meet the loads of today's IRC servers (around a thousand concurrent clients [18]).

## 7 Discussion

In this section we discuss some interesting advantages and challenges encountered during our development.

**The choreographic advantage**   The benefit of the choreographic programming style compared to writing the implementation of each role by hand can be seen by comparing the choreography with its projections, which follow the traditional style of using send and receive actions. A single statement in the choreography might translate to multiple ones in its projections, so we had to write less code. Moreover, a Choral class projects a separate class for each of its roles, which results in fewer components as is visible by comparing the two sides of Figure 5.

Another important difference between our IRC choreography and traditional implementations is that it is easier to follow: interactions are determined by clearly recognisable method calls and it is easy to see how events at a role trigger the enqueueing of events at another role. The projected code, instead, requires understanding the side-effects of network actions at different locations. We can illustrate this difference in complexity by displaying the interactions between the different classes in the source choreography and the compiled code (Figure 7). The first diagram shows the common IRC connection registration sequence (mentioned previously in Figure 2), where the user inserts the NICK and USER commands at the client. When the first command is communicated to the server, this does not generate any new events but only a state update. This is because the server must also wait for the USER command before proceeding. When it is received, we can see that the server enqueues a series of events that should take place at a later time. These events are then processed similarly by the handler for server events, which triggers the communication of different commands (the IRC 'server welcome') from the server to the client.

The second diagram tracks the same flow but in the projected implementation. It is substantially more complicated, not least because we have twice as many interacting

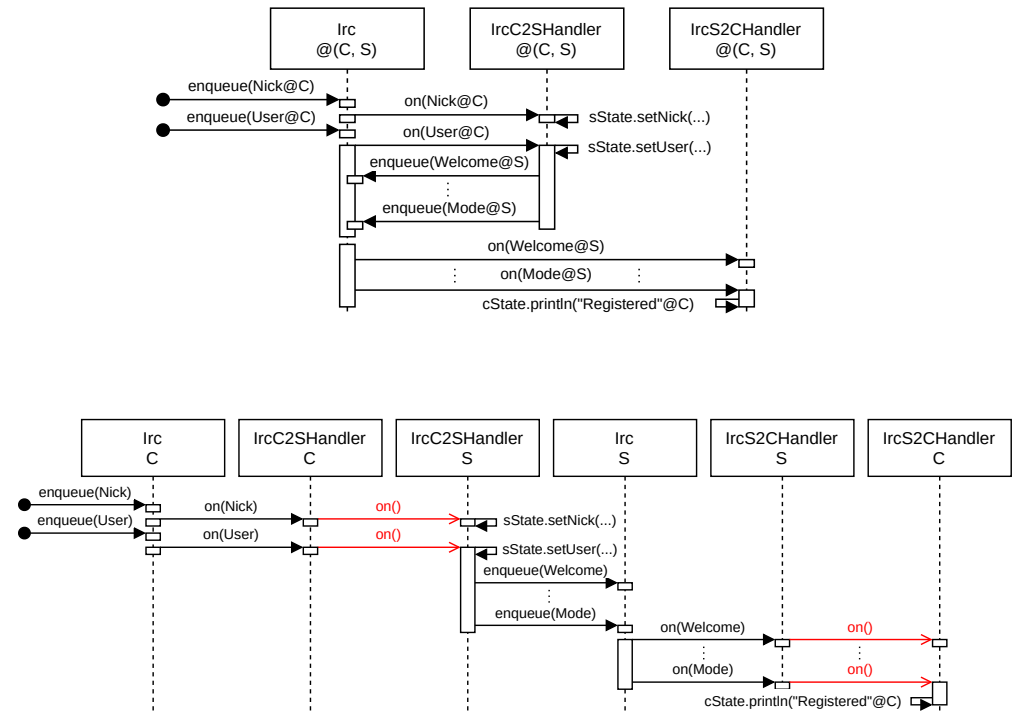

■ **Figure 7** Two interaction diagrams showcasing the difference between the choreographic (top) and the local (bottom) perspectives in the implementation of a protocol. We shorten some of the names for brevity.

classes. Furthermore, predicting what happens now requires the difficult analysis of how send and receive actions will match at runtime for message exchanges, indicated by red lines. In this specific case, sending actions are effectively asynchronous invocations of the remote side's projected method `on()`.

Our choreographic implementation comes with the same guarantees provided by any Choral program [19]. Notably, running the projections together cannot incur communication mismatches. This is because communications are implemented by invoking a single method, instead of separate send and receive actions. Also, when a role might send different types of messages, the compiler guarantees that the receiver will expect and react accordingly to any of these types. When interacting with third-party software, these guarantees depend on the correctness of such software.

Another benefit of adopting the choreographic style is readability. Since interactions are explicitly visible in the code, we felt that it was easier to check that we were implementing communications following the IRC specification, and that the corresponding computations were triggered correctly. In a sense, our implementation can be seen as (a first version of) a readable reference implementation of the IRC client–server protocol. By contrast, understanding what other third-party implementations do is more challenging, because their interactions are obtained as side-effects of running send and receive actions in separate programs at the right time (the usual challenge presented by concurrent code). Not having to write well-timed send and receive actions also saved time when we had to write code.

Overall, we believe that the benefits of the choreographic approach are relevant not only for the first implementation of a protocol but also for evolving such implementation with confidence in the future. Protocols are often evolved in order to offer new features, and choreographies help with producing and updating coherent implementations of all involved roles. In addition to helping developers, having a reference choreographic implementation of a protocol might aid designers. For example, they could edit a choreography to sketch modifications of a protocol and use the compiler to check if knowledge of choice is respected.

**Challenges**   Moving to the main challenge in writing our choreography, most of our development time was spent studying and modelling the full-duplex asynchronous interaction pattern exhibited by IRC. This is not very surprising considering that, so far, this pattern has not received much attention in the community of choreographic programming. Our implementation of this pattern has been written with modularity in mind, so future implementations of similar protocols can hopefully reuse (or at least be based on) our `Events` library. Nevertheless, we believe that our work showcases the need for more studies on the codification of reusable coordination patterns as choreographic programs. It is encouraging that Choral's higher-order choreographies and expressive type system allow for obtaining reusable libraries.

An additional minor technical challenge was the absence of language features that can help with speeding up code writing. In particular, Choral does not yet have syntax for inline lambda expressions or anonymous classes, making it necessary to create additional one-off named classes and choreographies. Some places in the code required the use of `synchronized` blocks in order to safeguard concurrent access to the state of the server that is shared between instances of the choreography. Unfortunately, Choral does not possess the `synchronized` keyword yet, but it was possible to get around the issue by calling out to Java code that acts as a thin wrapper. Finally, Choral requires one to annotate every type and data literal with a role. While useful in the case of types because it makes the choreography's data flow apparent, having to specify the role for data literals can get verbose, e.g., `counter + 1@A` or `"Name: "@A + name`. Judging from our experience, we believe that the code would be just as clear if the locations of literals were inferred from the context. None of these issues presented a roadblock in the development of the choreography, since they can be worked around in straightforward ways. Adding the necessary syntax and implementing a form of simple role inference for Choral would improve its usability, so we think it represents interesting future work.

## 8   Related Work

To the best of our knowledge, our work is the first that uses choreographic programming to develop an implementation of a real-world protocol which is capable of interoperating with existing third-party implementations.

The most closely related work to ours is the original presentation of Choral [19], the choreographic programming language that we used to develop our implementation of IRC. The original Choral paper introduces numerous examples, but none of them are aimed at our goal. Specifically, most are toy examples or present quite simplistic interaction patterns, neither of which are true of IRC. Furthermore, no attention is paid to the issue of interoperability with existing software, and as we have shown simple selections based on enumerations as in [19] are insufficient for that. By contrast, IRC has been used in the real world for decades with many different implementations and exhibits quite complex behaviour (full-duplex asynchronous communication).

Also related to our work are other implementations of choreographic programming languages. Chor [6, 36] and AIOCJ [15] both project their choreographies to the Jolie programming language [38], while Kalas [41] generates CakeML code. These works introduced different features, including session types for choreographies and explicit service endpoints, runtime modification of code, and a certified compilation pipeline. However, they do not support higher-order choreographies or allow for the possibility of defining custom communication semantics (channel implementations). Without those features, we would not have been able to achieve interoperability or encapsulate our full-duplex asynchronous `Events` pattern into a reusable component. HasChor [44] takes a different approach and implements choreographic programming as a domain-specific language embedded in Haskell. Choreographies in HasChor are values of a monadic type, which allows for using Haskell's higher-order composition but at the same time does not support the static analysis of choreographies. In particular, HasChor does not provide any control over selections, and any role evaluating a conditional must broadcast its decision to all other roles (even those not involved in the conditional). As we have discussed, having fine-grained control over which selections are made and how they are implemented is essential to our development.

A more weakly related direction of research is that on multitier programming [46]. Just like choreographic programming, multitier programs can talk about multiple roles. However, the way code is structured is very different: choreographic programming has a single 'objective' viewpoint that oversees all roles at once, which is what allows for writing explicitly how roles should interact; whereas multitier code is always written from the local 'subjective' viewpoint of a single role, which can be switched to that of another role by using appropriate primitives [21]. Thus multitier programs are essentially nested compositions of local code. Communications are not specified explicitly, but are rather left as implementation details to dedicated middlewares that can realise them as they wish. Having the full control over when communication should take place given by choreographic programming is important for writing faithful implementations of existing protocols, as in our case. Also, multitier programming languages typically come with fixed middlewares, but the feature of writing our own 'middleware' implementation (our IRC channels) is key to our development. The work in [45] is an exception, in that it allows at least for controlling the wire format.

Multiparty session types is another approach that leverages a global perspective on multiple roles [25], but it is less direct [23]. In this approach developers use 'global types' to express abstract protocols without computational details (e.g., internal checks and data serialisation cannot be captured). These are then projected to 'local types': local specifications of the allowed sequences of communication actions that a program can perform to implement a role. The programmer is then responsible for manually implementing these actions as expected. By contrast choreographic programming requires only the writing of the source choreography, which is then automatically compiled to an executable implementation.

Traditional approaches to distributed programming typically do not adopt a global view, e.g., actors in Akka [47] and Erlang [3], remote method invocations in Java RMI [40], and services in Jolie [38] and WS-BPEL [39]. In these approaches interactions are not syntactically manifest in programs, so they can be more error-prone [19, 31, 37]. In many of these approaches invoking an object method (or a service operation) can correspond to sending a message. Choral adds the possibility of having methods that implement both sides of a communication from a global viewpoint, which is the crucial feature for writing choreographic code.

In software engineering, providing architectural information about programs is an important topic, including architecturally-evident coding [17] and model-driven engineering approaches to microservices [20]. Choreographic programming can be seen as a way to achieve similar goals through the design of programming languages, and some choreographies can be seen as encodings of integration patterns [24].

## 9 Conclusion and Future Work

We have developed the first interoperable implementation of a real-world protocol (IRC) using choreographic programming. Our work demonstrates that it is possible to apply this emerging paradigm to the development of realistic software, and it reveals the practical relevance of higher-order choreographic composition and customisable communication semantics as in Choral.

From a theoretical viewpoint, our IRC choreography is essentially the parallel composition of two choreographies. Parallel choreographies are well-studied [5, 30, 43], but they require careful handling of shared state, such as the challenge that we encountered with concurrent usage of a bidirectional communication channel. While there are results on alternative composition operators that avoid this problem (e.g., choreographic choice [37]), they are still insufficiently expressive for full-duplex asynchrony. Our implementation motivates further investigations in this direction.

A future line of research opened by our work is, straightforwardly, the implementation of other real-world protocols based on different interaction patterns than the one found in IRC. Hopefully, these kinds of efforts can lead in the future to a collection of reusable choreographic libraries for interaction patterns that make it easy and quick to implement protocols of different kinds.

Another potential line of future work is making the Choral compiler more extensible. In our case, we had to modify the compiler to deal with our new form of selection. However, more general or different forms of selections or other primitives might be needed to capture other protocols. To this end, allowing for user-defined extensions of the Choral compiler (and choreographic languages in general) could be of help.

**Acknowledgements**    Work partially supported by Villum Fonden, grants no. 29518 and 50079, and the Independent Research Fund Denmark, grant no. 0135-00219.

## A  Conformance Tests

In order to use `irctest` we first start the server and instruct it to use `My.Little.Server` as its hostname by passing it as the first command-line argument. This is a minor detail necessary only because the suite hardcodes the hostname expected in the server's replies.

Once the server is up and running, we invoke the suite in the following way: the environment variables `IRCTEST_SERVER_HOSTNAME` and `IRCTEST_SERVER_PORT` specify the address and the port of the server, while the `-k` parameter is used to provide a boolean expression that selects which tests to run. We only select the tests corresponding to basic functionality and make sure to deselect more modern features that are not part of the original specifications and that we do not implement, such as IRCv3, message tags, etc.

```
IRCTEST_SERVER_HOSTNAME=127.0.0.1 IRCTEST_SERVER_PORT=8667 \
  pytest --controller irctest.controllers.external_server \
    -k "$(cat <<EOF
      not IRCv3 and \
      not labeled-response and \
      not account-tag and \
      not message-tags and \
      ((testNick or \
        testStarNick or \
        testEmptyNick or \
        testFailedNickChange or \
        testEarlyNickCollision or \
        testEmptyRealname) or \
      testPing or \
      testJoin or \
      (testPart or testBasicPart) or \
      testPrivmsg or \
      testLineAtLimit or \
      (testQuit and not Ergo))
EOF
)"
```

The generated test report shows that we pass all of the selected tests and that our IRC server implementation conforms to the standard IRC behaviour:

```
===== test session starts =====
platform linux -- Python 3.11.2, pytest-7.2.1, pluggy-1.0.0
rootdir: /home/irc/progval-irctest, configfile: pytest.ini
collected 435 items / 389 deselected / 46 selected

irctest/server_tests/connection_registration.py .....   [ 17%]
irctest/server_tests/join.py ....                       [ 31%]
```

```
irctest/server_tests/messages.py .....          [ 48%]
irctest/server_tests/part.py .....              [ 65%]
irctest/server_tests/pingpong.py ...            [ 75%]
irctest/server_tests/quit.py .                  [ 79%]
irctest/server_tests/regressions.py ......      [100%]
===== 29 passed, 389 deselected, 435 warnings in 5.76s =====
```

A full list of tests that have been run can be gathered by passing the `--collect-only` option to `pytest`:

```
<Package server_tests>
  <Module connection_registration.py>
    <Class ConnectionRegistrationTestCase>
      <Function testQuitDisconnects>
      <Function testQuitErrors>
      <Function testNickCollision>
      <Function testEarlyNickCollision>
      <Function testEmptyRealname>
  <Module join.py>
    <Class JoinTestCase>
      <Function testJoinAllMessages>
      <Function testJoinNamreply>
      <Function testJoinTwice>
      <Function testJoinPartiallyInvalid>
  <Module messages.py>
    <Class PrivmsgTestCase>
      <Function testPrivmsg>
      <Function testPrivmsgNonexistentChannel>
      <Function testPrivmsgToUser>
      <Function testPrivmsgNonexistentUser>
    <Class LengthLimitTestCase>
      <Function testLineAtLimit>
  <Module part.py>
    <Class PartTestCase>
      <Function testPartNotInEmptyChannel>
      <Function testPartNotInNonEmptyChannel>
      <Function testBasicPart>
      <Function testBasicPartRfc2812>
      <Function testPartMessage>
  <Module pingpong.py>
    <Class PingPongTestCase>
      <Function testPing>
      <Function testPingNoToken>
      <Function testPingEmptyToken>
  <Module quit.py>
    <Class ChannelQuitTestCase>
```

```
    <Function testQuit>
 <Module regressions.py>
   <Class RegressionsTestCase>
     <Function testFailedNickChange>
     <Function testStarNick>
     <Function testEmptyNick>
     <Function testNickRelease>
     <Function testNickReleaseQuit>
     <Function testNickReleaseUnregistered>
```

## About the authors

**Lovro Lugović** is a PhD student at the University of Southern Denmark under the supervision of Fabrizio Montesi. His research interests focus on programming language design and concurrency theory. Contact him at lugovic@imada.sdu.dk.
https://orcid.org/0000-0001-9684-9567

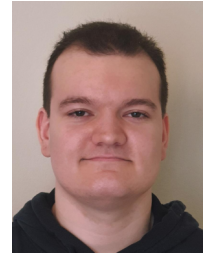

**Fabrizio Montesi** is Professor of Computer Science at the University of Southern Denmark. He is a Villum Young Investigator and recipient of several awards for science and innovation, including the EAPLS Best PhD Dissertation Award and the Best Thesis in ICT Award from the General Confederation of Italian Industry. Contact him at fmontesi@imada.sdu.dk.
https://orcid.org/0000-0003-4666-901X

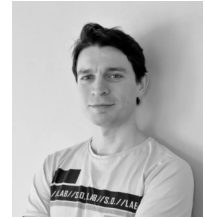